\documentclass{svjour3}
\usepackage{amsmath}
\usepackage{amsfonts}
\usepackage{amssymb}
\usepackage{graphicx}
\usepackage{float}
\usepackage[dvipsnames]{xcolor}
\usepackage[english,frenchb]{babel}
\usepackage{multirow}
\usepackage[square,sort,comma,numbers]{natbib}
\usepackage{pifont}
\usepackage[scale=0.8]{geometry}

\begin{document}
\title{Conditional averaging on volumetric velocity fields for analysis of the pseudo-periodic organization of jet-in-crossflow vortices}

\titlerunning{Conditional averaging for analysis of the vortices of a jet in crossflow} 

\author{Tristan CAMBONIE}

\institute{T. CAMBONIE \at
              Laboratoire Physique et M\'ecanique des Milieux H\'et\'erog\`enes (PMMH), UMR 7636
ESPCI\\ 10, rue Vauquelin - 75231 Paris Cedex 5
 \\
              Tel.: +33(0) 1 69 15 80 86\\
              \email{tristan.cambonie@espci.fr}}

\date{Received: date / Accepted: date}
\maketitle

\begin{abstract}
Volumetric velocimetry measurements have been performed on a round jet in crossflow (JICF). Two kind of conditional averaging process are presented : a global conditional average (GCA) and a local conditional average (LCA). Vortices crossing the JICF symmetry plane are used to quantify the jet pseudo-periodicity leading to the GCA and LCA definitions. Because they make possible to improve the velocity field resolution as well as to significantly reduce the experimental noise, these conditional averages are an interesting and efficient way to study the instantaneous swirling structures of this instantaneous flow.
\end{abstract}

\section{Introduction}
This last decade, with the development of volumetric velocimetry techniques it became possible to directly visualize and study the instantaneous 3D structures in experimental flow fields. Nevertheless, the resolutions of the experimental instantaneous 3D velocity fields are still low (1 to 4 vectors per mm) in comparison of what can be achieved numerically. Indeed, the experimental resolutions depends strongly on the tracer concentration in the flow (bubbles, particles...), concentration which is itself limited by the visual screening phenomenon \citep{CaAi14} which occurs between the tracers. As a result, it limits the size of the structures which can be observed.
\begin{figure}[htbp!]
\begin{center}
\begin{tabular}{c}
	\includegraphics[width=0.7\textwidth]{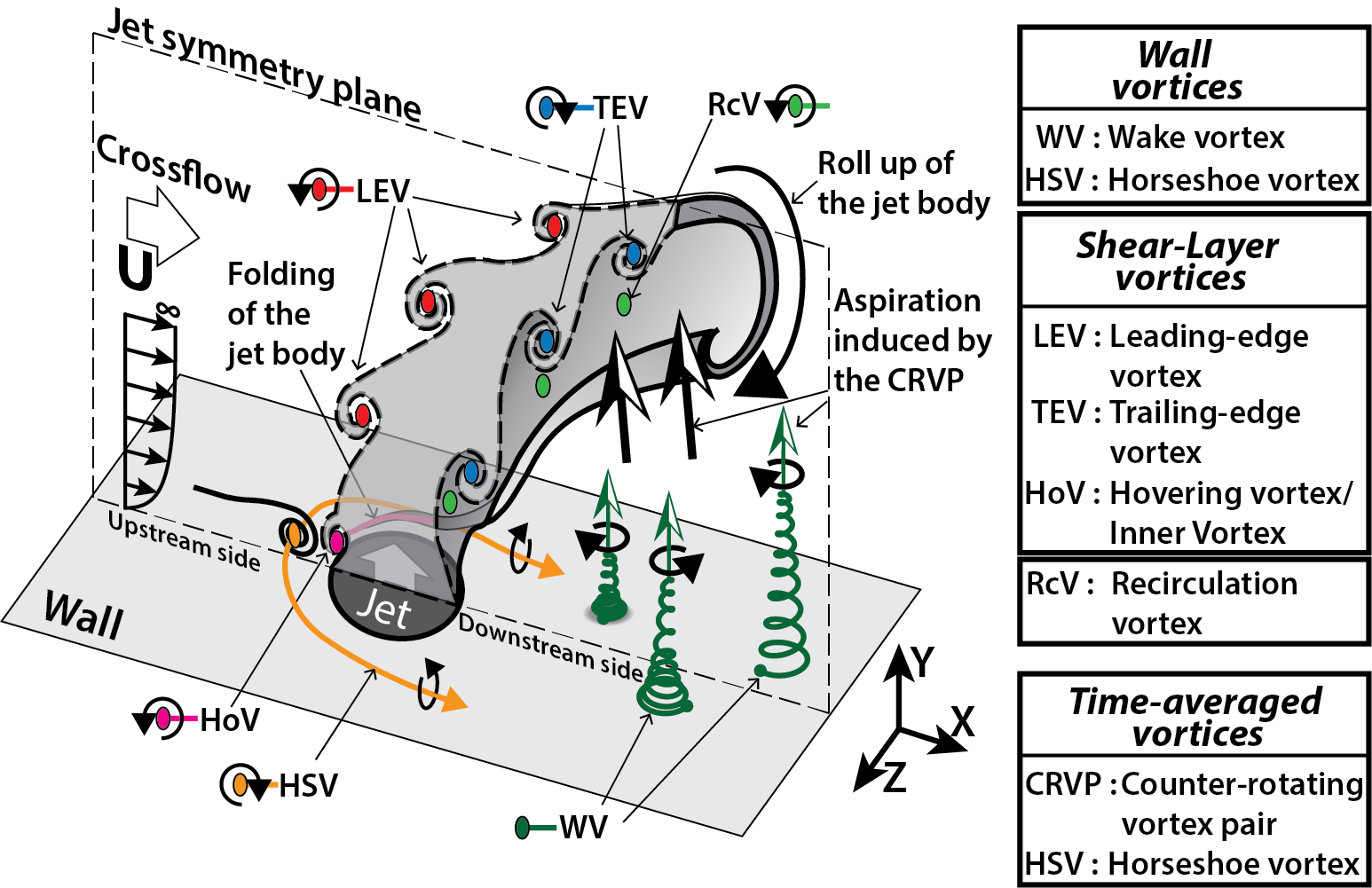}
\end{tabular}
\caption{JICF instantaneous topology \citep{CamAi2014}.}
\label{fig:intro}
\end{center}
\end{figure}
For periodic flows, phase averaging is an adequate solution to increase the resolution of the velocity fields and extract less noisy instantaneous swirling structures from it. Nevertheless, such approach is harder to implement in the case of pseudo-periodic flows or in the case of flows whose periodicity is not controlled by the experimentalist. 
The Jet In CrossFlow (JICF) is a flow configuration involved in many industrial processes such as film cooling, fuel injection, polluant dispersion or flow control...
The JICF is a junction flow between a jet and a transverse boundary-layer crossflow (Fig. \ref{fig:intro}) whose interaction forms a complex intricated system of vortices, including the upstream and downstream shear layer vortices (resp. Leading-edge and Trailing-edge vortices in Fig. \ref{fig:intro}). Their dynamic is complex and pseudo-periodic. A recent review on the topic can be found in \citep{CamAi2014}. A better and deep understanding of the physic of these instantaneous vortical structures is essential for most of the aforementionned applications of jets in crossflow. 
In this study, two kind of conditional averaging process are then presented. They are based on the locations of the vortices in the symmetry plane as well as their statistical properties in order to  improve the visualization of the instantaneous swirling structures of the JICF.%
\section{Experimental setup, volumetric velocimetry method, and vortex detection criterion}
\subsection{Experimental setup}
\begin{figure}[htbp!]
\begin{center}
\begin{tabular}{cc}
\includegraphics[width=0.45\textwidth]{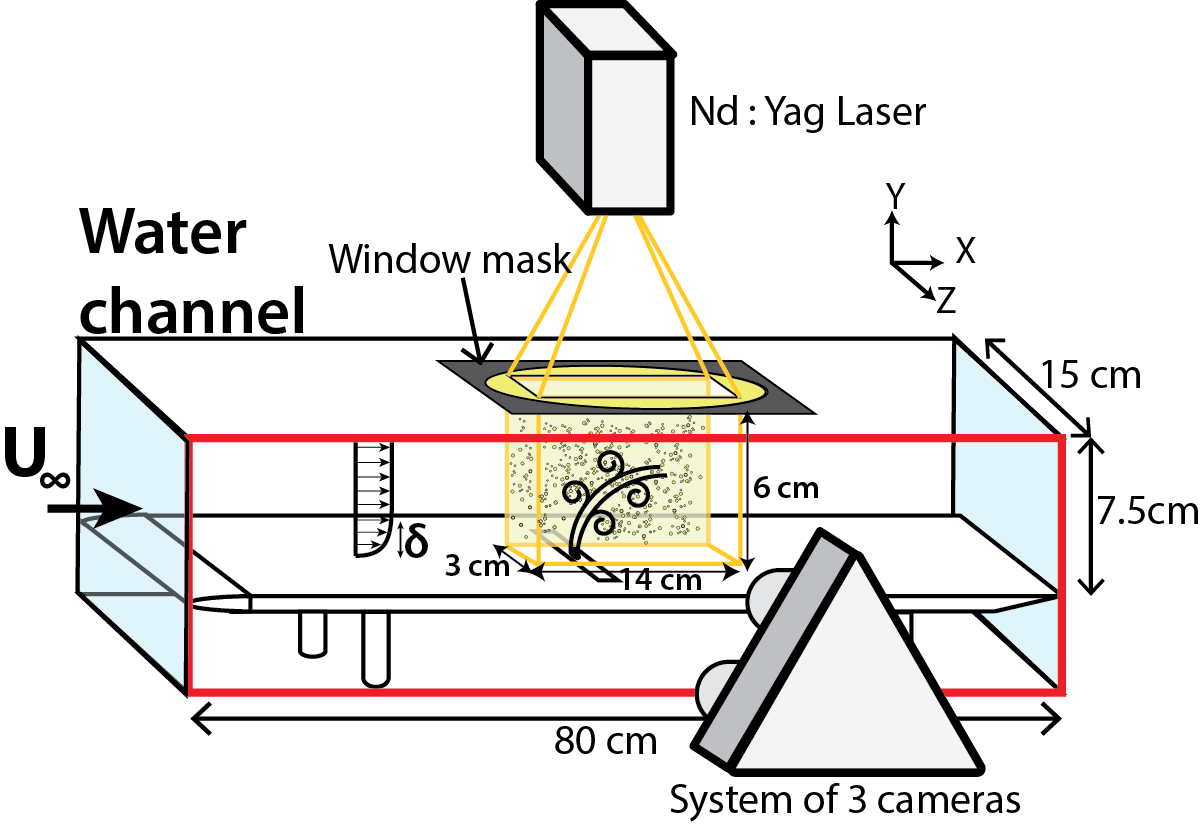}
&\includegraphics[width=0.27\textwidth]{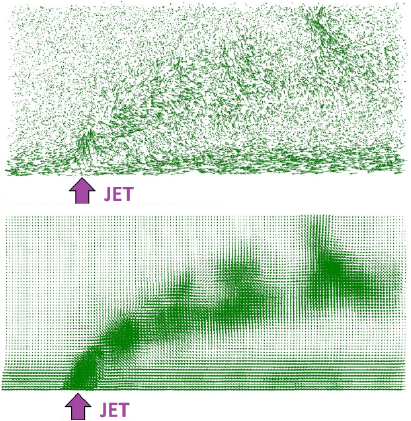}\\
a)&b)\\
\end{tabular}
\caption{a) Experimental setup. b) Top : Side view of an instantaneous 3D raw velocity field. Bottom : Side view of the same 3D velocity field after the interpolation step. To improve the visualization of the velocity fluctuations, $U_{\inf}$ has been substracted from the data.}
\label{fig:expSetup}
\end{center}
\end{figure}

Experiments have been performed in a gravity-driven hydrodynamical channel on a round transverse jet at low velocity ratio (Fig. \ref{fig:expSetup}a). 
The jet is also gravity-driven. Its water tank is filled with the same water used in the channel and has therefore the same particle concentration than the rest of the channel. The round jet exit is flush with respect to the flat plate and its geometrical center coincides with the axis system origin. The geometrical and physical parameters of the experiment presented in this article are the following : the bulk channel velocity $U_{\inf}=9\ cm.s^{-1}$, the mean jet velocity over the jet exit $V_{jet}=15\ cm.s^{-1}$, the velocity ratio $VR=V_{jet}/U_{\inf}=1.67$, the jet diameter $d_{jet}=8\ mm$, the boundary-layer thickness $\delta=15\ mm$, the crossflow Reynolds number $Re_{cf}=U_{\inf}\cdot\delta/\nu=1350$, the jet Reynolds number $Re_{jet}=V_{jet}\cdot d_{jet}/\nu=1200$.
The flow is seeded with 50 $\mu$m particles, with a visual concentration of $4.5\times10^{-2}$ particles per pixel\citep{CaAi14}. 
The flow is illuminated through the upper wall and the particles are tracked in the volume using three cameras facing the side wall. The three cameras of this system are 4~MP double-framed with a 12~bit output. Volumetric illumination is generated using a 200~mJ pulsed Nd:YAG laser and two perpendicular cylindrical lenses. Synchronization is ensured by a TSI synchronizer. The measurement volume is 14$\times$6$\times$3~cm$^3$ and is homogeneously illuminated. 1000 instantaneous velocity fields have been recorded.
\subsection{Volumetric 3-component Velocimetry}
3D defocusing digital particle image velocimetry (3D DDPIV) measurements have been performed using a  system using three cameras (4~MP, double-framed, 12~bit output) designed by TSI (Volumetric 3-component Velocimetry system, V3V) on the basis of the work of \citet{Pereira2002}. In a first stage, the intensity peaks corresponding to each particles are detected in each camera frame for each time step. Then, using a spatial calibration, the triplets of 2D particle coordinates are used to reconstruct for each time step a 3D field of particle positions. A particle tracking step, between $t$ and $t+\delta t$, leads to the instantaneous raw velocity field (Fig. \ref{fig:expSetup}b top). Finally, a last step interpolates this raw velocity field on a grid (Fig. \ref{fig:expSetup}b bottom), in order to be able to use classical visualization tools and more generally to post-process the data. More details can be found in\citep{CaAi14,Pereira2002,Cambonie2013}. Using the work of Cambonie\citep{CaAi14}, the set-up was designed and the physical (particle concentration, measurement volume dimensions...) and numerical parameters were chosen to optimize the quality and resolution of the instantaneous velocity fields while avoiding the optical screening of the particles in the back of the measurement volume by the  particles in the front layers. Indeed, a relationship exists between the imaged concentration and an optimal final spatial resolution of the interpolated instantaneous velocity fields. This optimal resolution is defined as the smaller resolution of the interpolation grid where the interpolation is adequately and successfully performed for more than 99\% of the voxels. 
\subsection{Vortex detection criterion}
	Instantaneous and mean swirling structures of the flow are visualized using isosurfaces of $\lambda_{ci}$. This vortex detection criterion has been initially proposed by \citet{Zhou1999} and improved by \citet{Chakraborty2005,Chakraborty2007}, and corresponds to the imaginary part of the complex eigenvalues of the diagonalized gradient velocity tensor $\mathcal{D}=\overrightarrow{\nabla}\overrightarrow{U}$. It measures the rotation frequency of a fluid particle around the vortex core in the main strain direction. 
The same definition is applied to the 2D gradient velocity tensor $\mathcal{D_{XY}}$ (resp. $\mathcal{D_{YZ}}$ and $\mathcal{D_{XZ}}$) in the XY plane (resp. YZ and XZ)  to define a similar criterion $\lambda_{ci\ Z}$ (resp. $\lambda_{ci\ X}$ and $\lambda_{ci\ Y}$) which only detect swirling motions along the Z direction (resp. X and Y directions).

\section{Global Conditional Average (GCA) }
\subsection{Vortex topology in the symmetry plane}
\begin{figure}[htbp!]
\begin{center}
\begin{tabular}{cc}
	\includegraphics[width=0.4\textwidth]{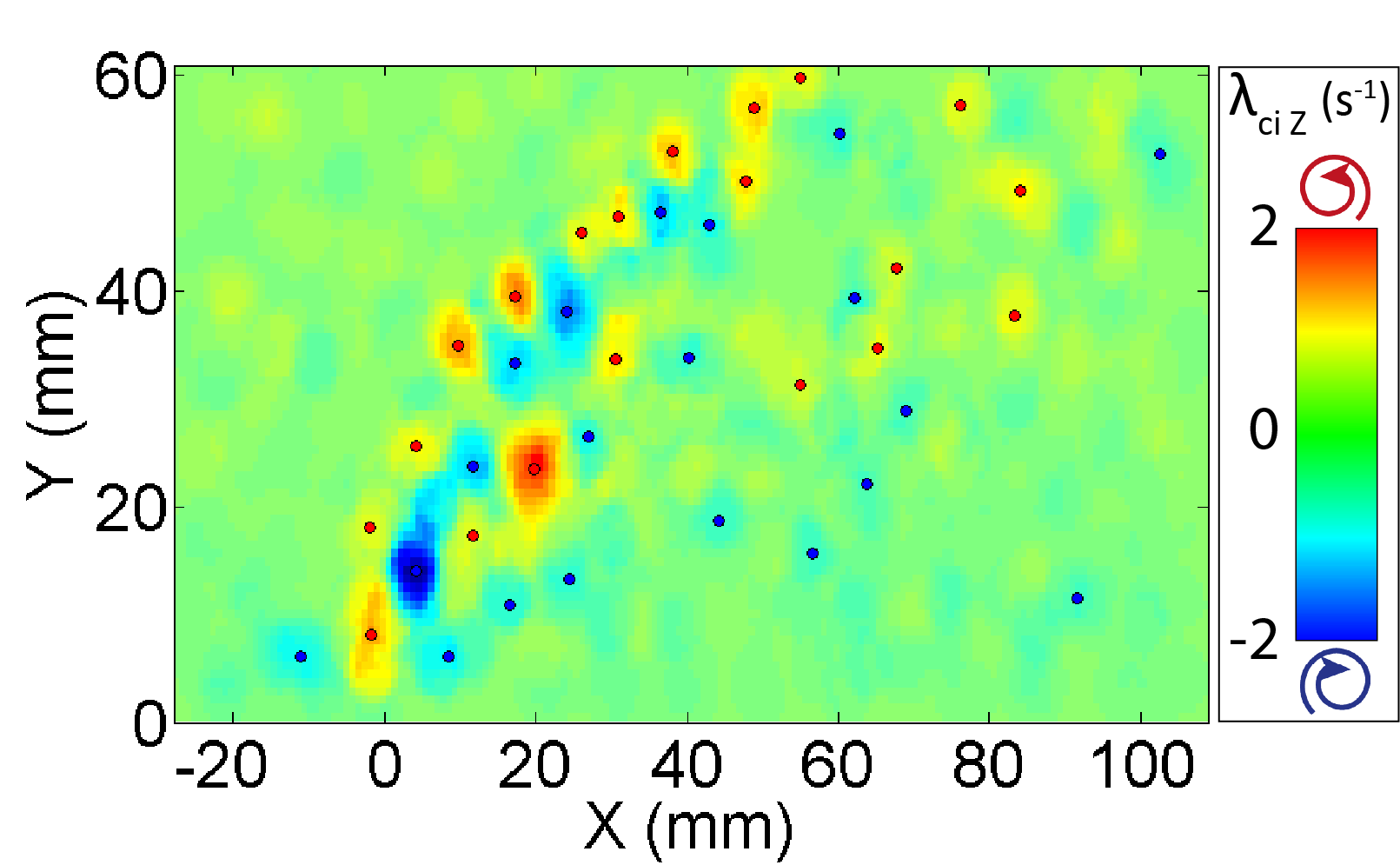} & 
	\includegraphics[width=0.395\textwidth]{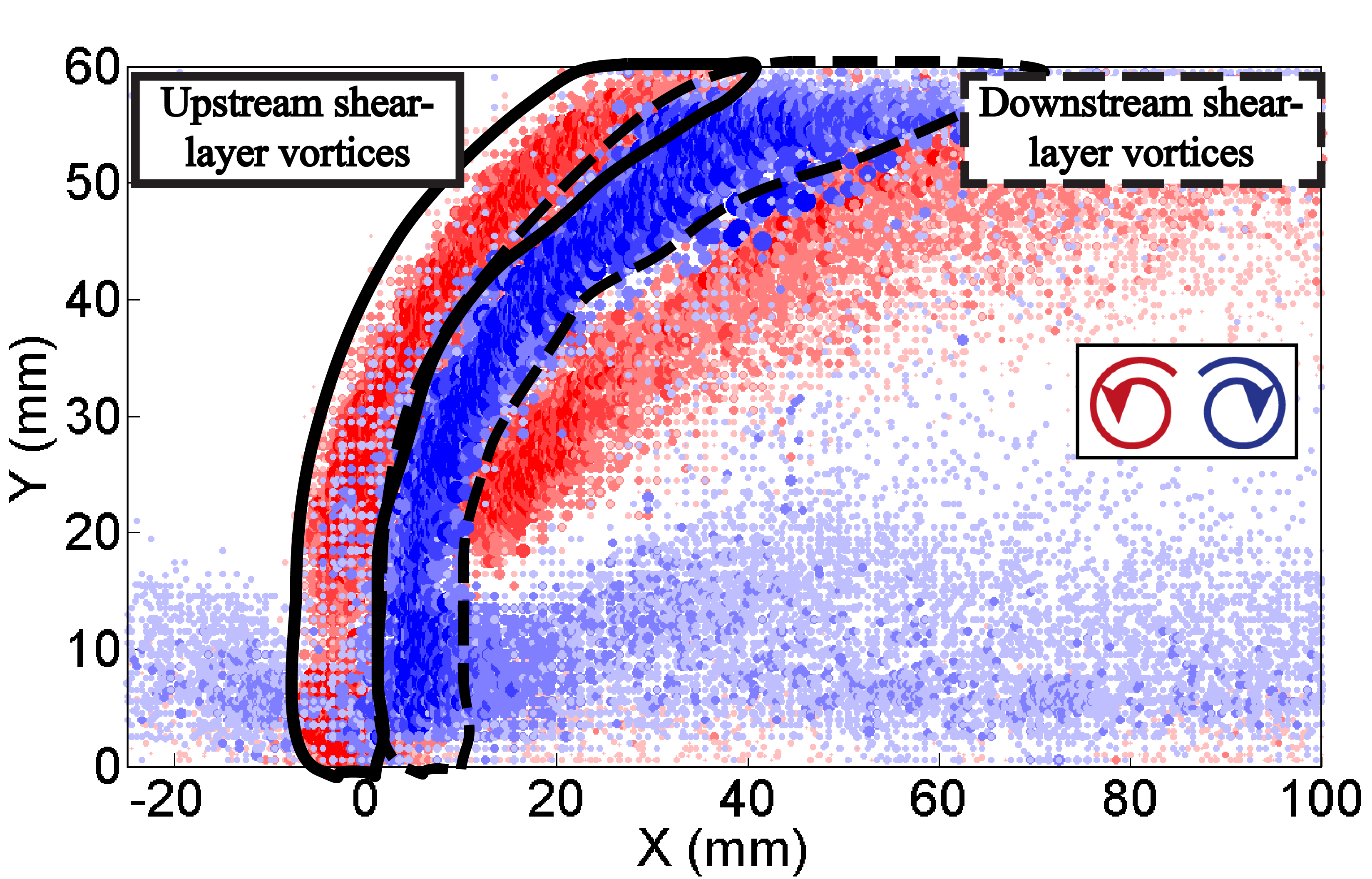}\\
a)&b)
\end{tabular}
\caption{a) Detection of the positive and negative maxima of $\lambda_{ci\ Z}$ in the jet symmetry plane. b) Cumulative distribution of the positive and negative vortices detected in 1000 velocity fields. The size and color intensity of each marker are proportionnal to the swirling intensity $\lambda_{ci\ Z}$  of each vortex.}
\label{fig:BlobDetect}
\end{center}
\end{figure}
Because it crosses most of the vortical structures of the JICF, the symmetry plane (Fig. \ref{fig:intro}) is a privileged plane to study the JICF pseudo-periodicity. 

Local positive and negative $\lambda_{ci\ Z}$ maxima are detected in the symmetry plane (Fig. \ref{fig:BlobDetect}a). A second step refined the maxima neighborhoods to detect a sub-grid position of each maximum. This detection is carried out for all 1000 velocity fields. All the positions of the vortices detected in the symmetry plane are put and displayed together (Fig. \ref{fig:BlobDetect}b).
The cumulative distribution of the vortex positions has distinct areas linked with the JICF topology \citep{CamAi2014}. 
In this study, we focus on the downstream shear-layer vortices (Fig. \ref{fig:BlobDetect}b, also called trailing-edge vortices in Fig. \ref{fig:intro}), but the process is the same for the upstream shear-layer vortices.
\subsection{Features of the downstream shear-layer vortices}
\subsubsection{Mean trajectory}
\begin{figure}[htbp!]
\begin{center}
\hspace*{-0.7cm}\begin{tabular}{cccc}
	\includegraphics[width=0.25\textwidth]{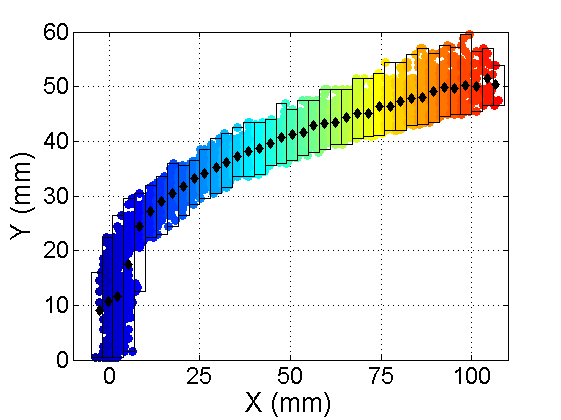}&
	\includegraphics[width=0.25\textwidth]{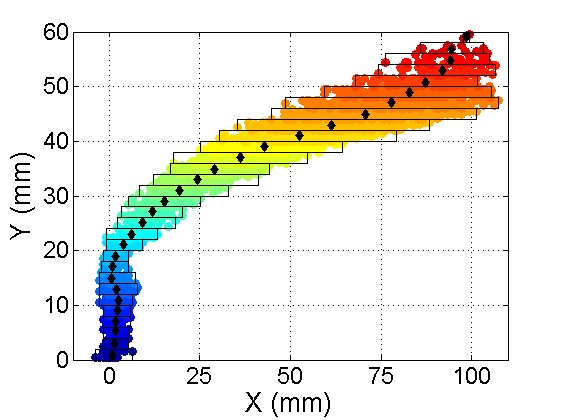}&
	\includegraphics[width=0.25\textwidth]{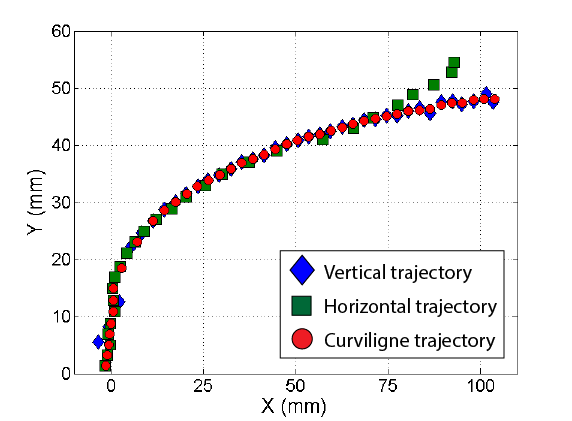} & 
	\includegraphics[width=0.25\textwidth]{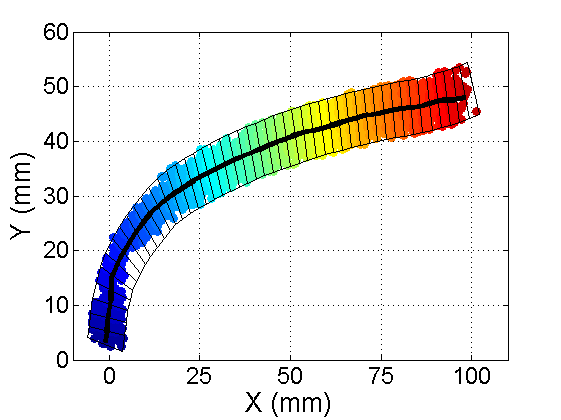}\\
a)&b)&c)&d)
\end{tabular}
\caption{a) Vertical slices. b) Horizontal slices. c) Vertical, horizontal and curvilinear trajectories. d) Oblique slices perpendicular to the rough curvilinear trajectory.}
\label{fig:trajTranche}
\end{center}
\end{figure}
To generate a mean trajectory, the cloud of downstream shear-layer vortices is first cut into vertical (Fig. \ref{fig:trajTranche}a) and horizontal (Fig. \ref{fig:trajTranche}b) slices. In each slice, the barycentre of the vortex positions pondered by the swirling intensity of each vortex is computed. The sets of mean vortex positions form different trajectories depending on whether the slices are vertical or horizontal. These two trajectories are then used to define a rough curvilinear trajectory (Fig. \ref{fig:trajTranche}c), which takes into account the direction changes of the jet. Finally, oblique slices are made in the vortex cloud perpendicularly to the rough curvilinear trajectory (Fig. \ref{fig:trajTranche}d), in which mean vortex positions (still pondered by the swirling intensity) are computed. They form a fine curvilinear trajectory along which is defined the curvilinear mean abscissa $S$ of the downstream shear-layer vortex cloud.
\subsubsection{Inter-vortex distance along the shear-layer trajectory}
\begin{figure}[htbp!]
\begin{center}
\begin{tabular}{cc}
	\includegraphics[width=0.4\textwidth,height=0.3\textwidth]{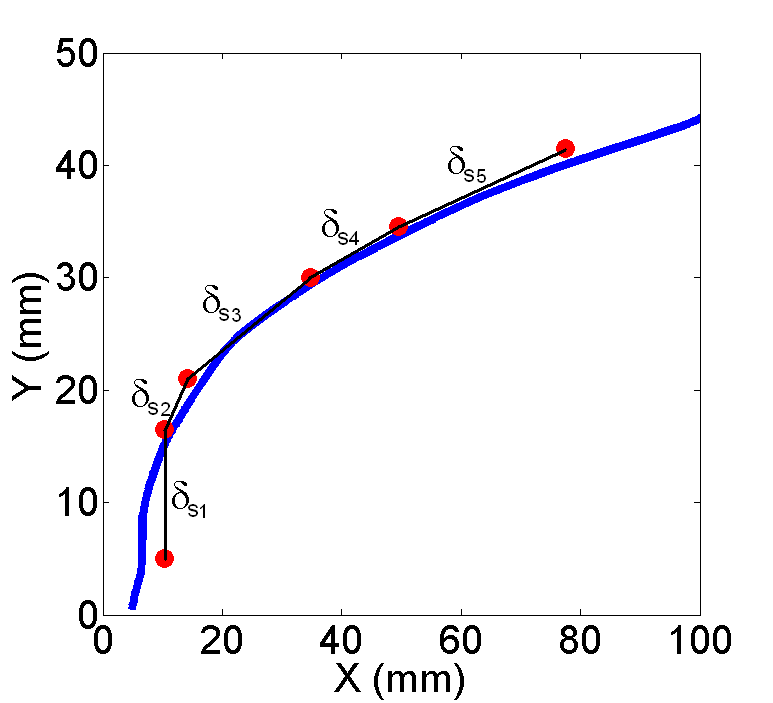} & 
	\includegraphics[width=0.4\textwidth,height=0.3\textwidth]{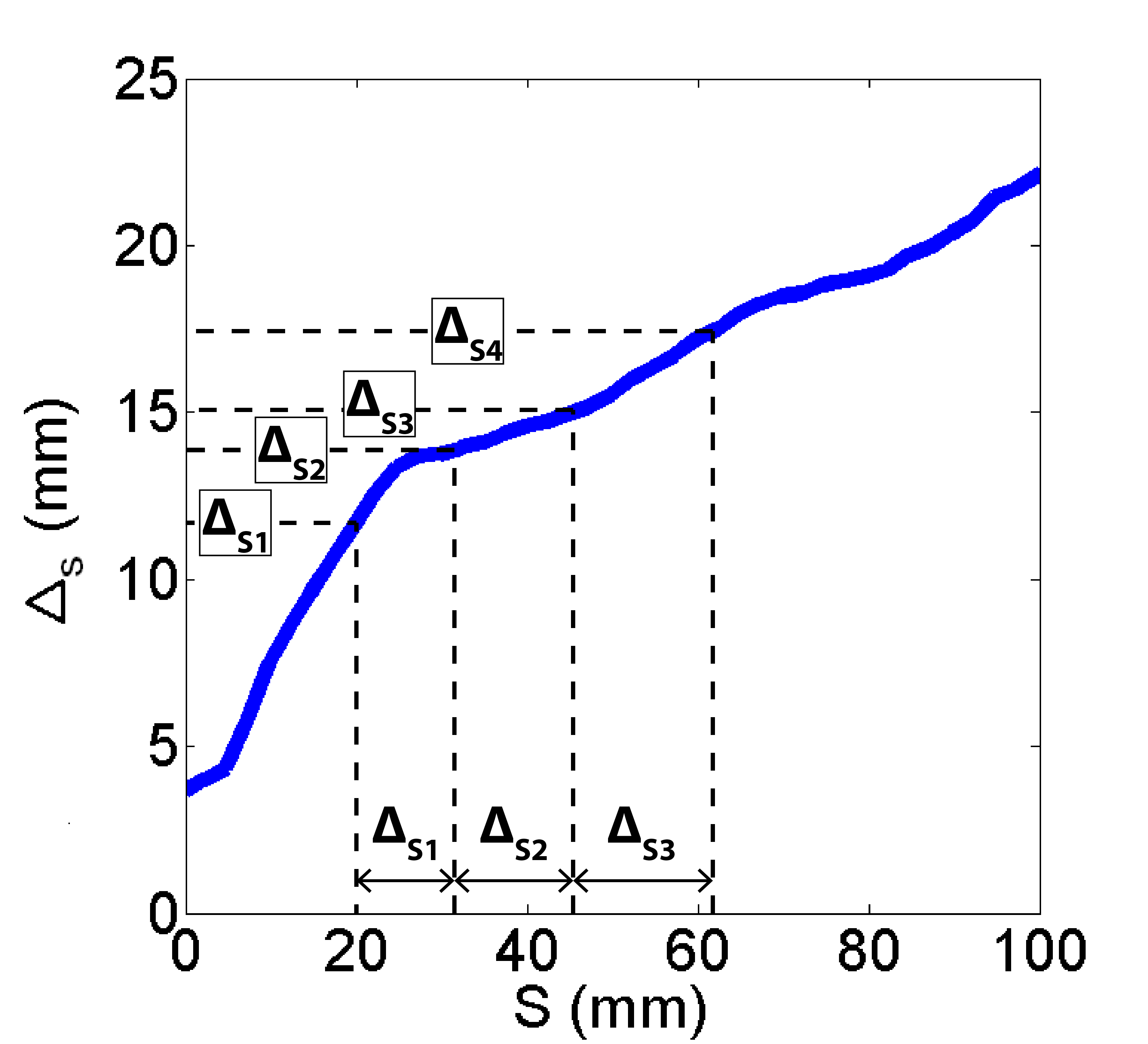}\\
a)&b)
\end{tabular}
\caption{a) Example of inter-vortex distances $\delta_{Si}$ between consecutive vortices (\textcolor{red}{$\circ$}) along the downstream shear-layer mean trajectory. b) Mean inter-vortex distance $\Delta_{S}$ along the downstream shear-layer mean abscissa $S$.}
\label{fig:Traj+deltaP}
\end{center}
\end{figure}
In each velocity field, the distance $\delta_S$ between two consecutive vortices is measured (an example is given in Fig. \ref{fig:Traj+deltaP}a) and is associated with the position of the first vortex along the curvilinear abscissa. Combining this information over 1000 time steps, we get the mean inter-vortex distance $\Delta_S$ along the shear-layer trajectory (Fig. \ref{fig:Traj+deltaP}b). It means that if along the curvilinear abscissa a vortex is located at the position S, the most likely position (statistically speaking) of the next vortex along this abscissa is $S+\Delta_S$. Fig. \ref{fig:Traj+deltaP}b shows $\Delta_S$ as a function of $S$. By instance, the vortex following the vortex located at S=20 mm is most likely to be located around $S=20+\Delta_{S=20mm}=20+12.4=32.4\ mm$, then the next one will be around $S=32.4+\Delta_{S=32.4mm}=32.4+14.1=46.5\ mm$ etc.
\subsection{Global conditional averaging method}
\begin{figure}[htbp!]
\begin{center}
\begin{tabular}{cc}
	\includegraphics[width=0.41\textwidth]{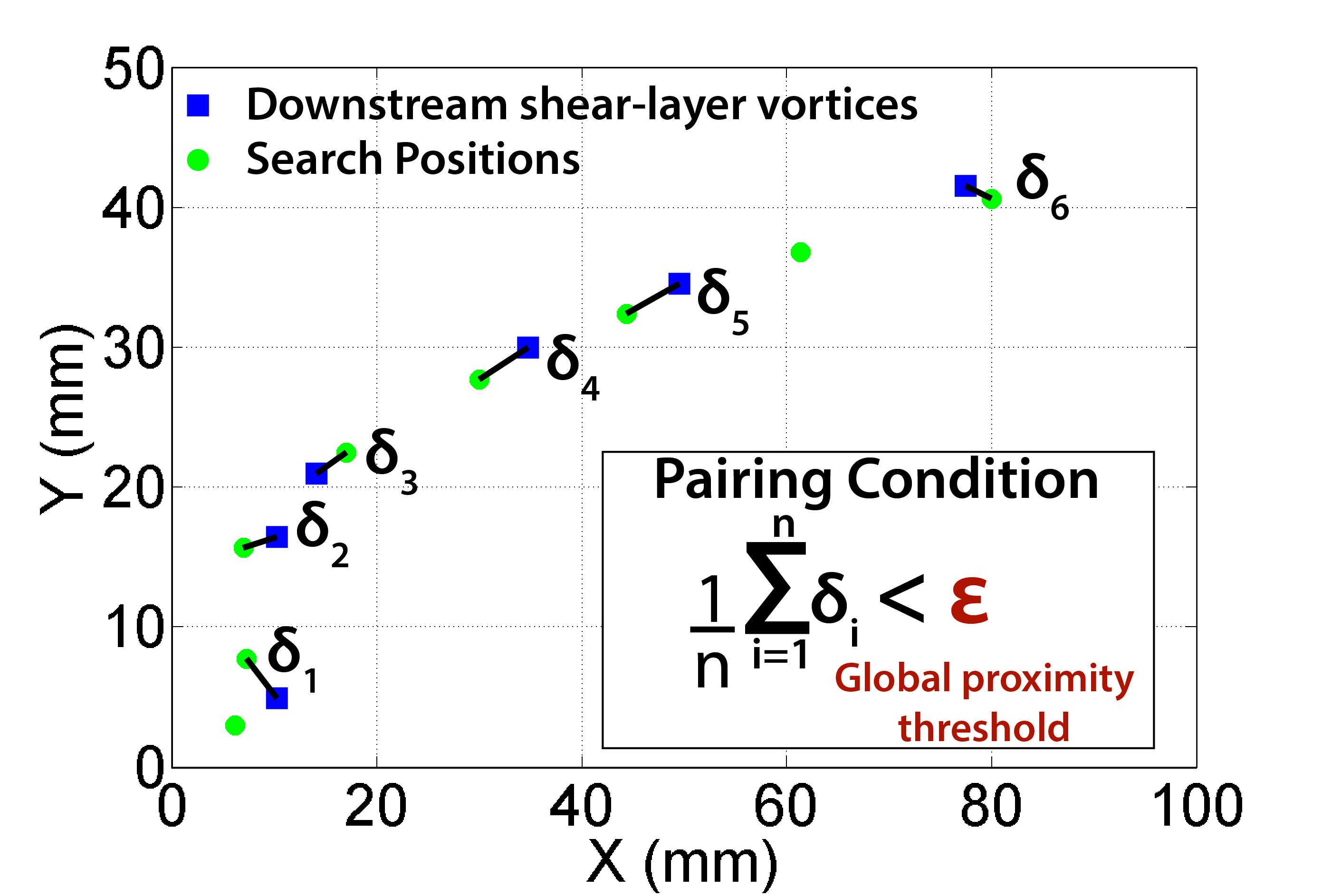} & 
	\includegraphics[width=0.4\textwidth]{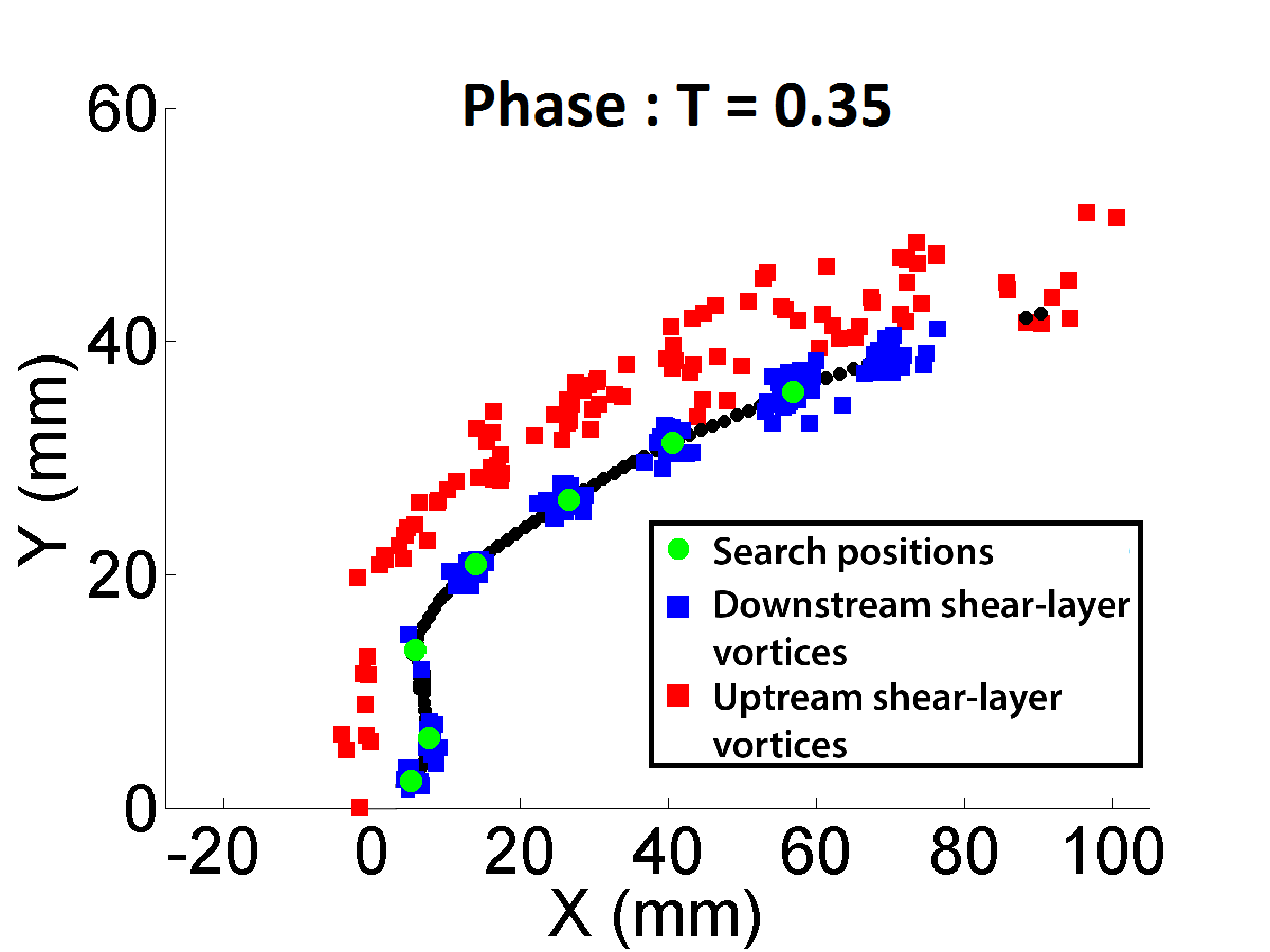}\\
a)&b)
\end{tabular}
\caption{a) Pairing between the instantaneous downstream shear-layer vortices in the symmetry plane (\textcolor{blue}{$\square$}) and the closest search positions (\textcolor{green}{$\circ$}) associated with normalized phase T=0.35. b) For the same phase, reference search positions, upstream and downstream shear-layer vortices.}
\label{fig:MCG-construction}
\end{center}
\end{figure}
Each instantaneous velocity field is evaluated and associated with a specific phase. 
First the shear-layer trajectory is coupled with the evolution of the mean inter-vortex distance $\Delta_S$ along this trajectory  to generate for each phase a set of expected positions along the shear-layer trajectory. By instance, for the normalized phase T=0.35, these expected positions are displayed using green circles in Fig. \ref{fig:MCG-construction}a. These positions are different for each phase and are used as reference positions to associate each instantaneous velocity field with its right phase. Each vortex of a given instantaneous velocity field is linked with the closest reference position (Fig. \ref{fig:MCG-construction}a). The average distance between the vortices and the reference positions is used as a norm to quantify the proximity of the instantaneous field with the phase associated with these reference positions. Among all the phases, the one with the smallest average distance is associated with this instantaneous field if this average is below a proximity threshold $\epsilon$ ($\epsilon$=2 mm for the GCA in Fig. \ref{fig:MCG-construction}a). This latter condition makes sure the vortex of the evaluated instantaneous field are close enough from the reference positions. Hence, an instantaneous field can be matched with no phase. This process is then repeated for each instantaneous field. After this step, each phase is associated with multiple instantaneous velocity fields from which an average is computed. For the normalized phase T=0.35, Fig. \ref{fig:MCG-construction}b shows the superposition of all the shear-layer vortices found in the instantaneous fields associated with this phase. As expected, the vortices of the downstream shear-layer are well-grouped in the neighborhood of the reference positions. On the opposite, the upstream shear-layer vortices does not show the same organization. It is a good illustration of the jet pseudo-periodicity. Even if upstream and downstream shear layer vortices may appear synchronized (Fig. \ref{fig:BlobDetect}a), the flow conditions differ on both side of the jet. It leads to slightly different shear-layer characteristics. As a result, the generation rate of the shear-layer vortices is different on the upstream and downstream shear-layers. The global conditional average (GCA) has therefore to be computed independently for both shear-layer.\\
Finally for each phase the raw velocity fields (one example of them is shown in Fig. \ref{fig:expSetup}b top) associated with this phase are concatenated into a single dense raw velocity field which is then interpolated on a regular tridimensionnal grid. Figure \ref{fig:MCGResults}a,b shows the upstream and downstream shear-layer with isosurfaces of $\lambda_{ci\ Z}$. The GCA is displayed in Fig. \ref{fig:MCGResults}b while Fig \ref{fig:MCGResults}a shows one of the instantaneous fields associated with the GCA.
\subsection{Results and discussion on GCA}
\begin{figure}[htbp!]
\begin{center}
\begin{tabular}{cc}
	\includegraphics[width=0.4\textwidth]{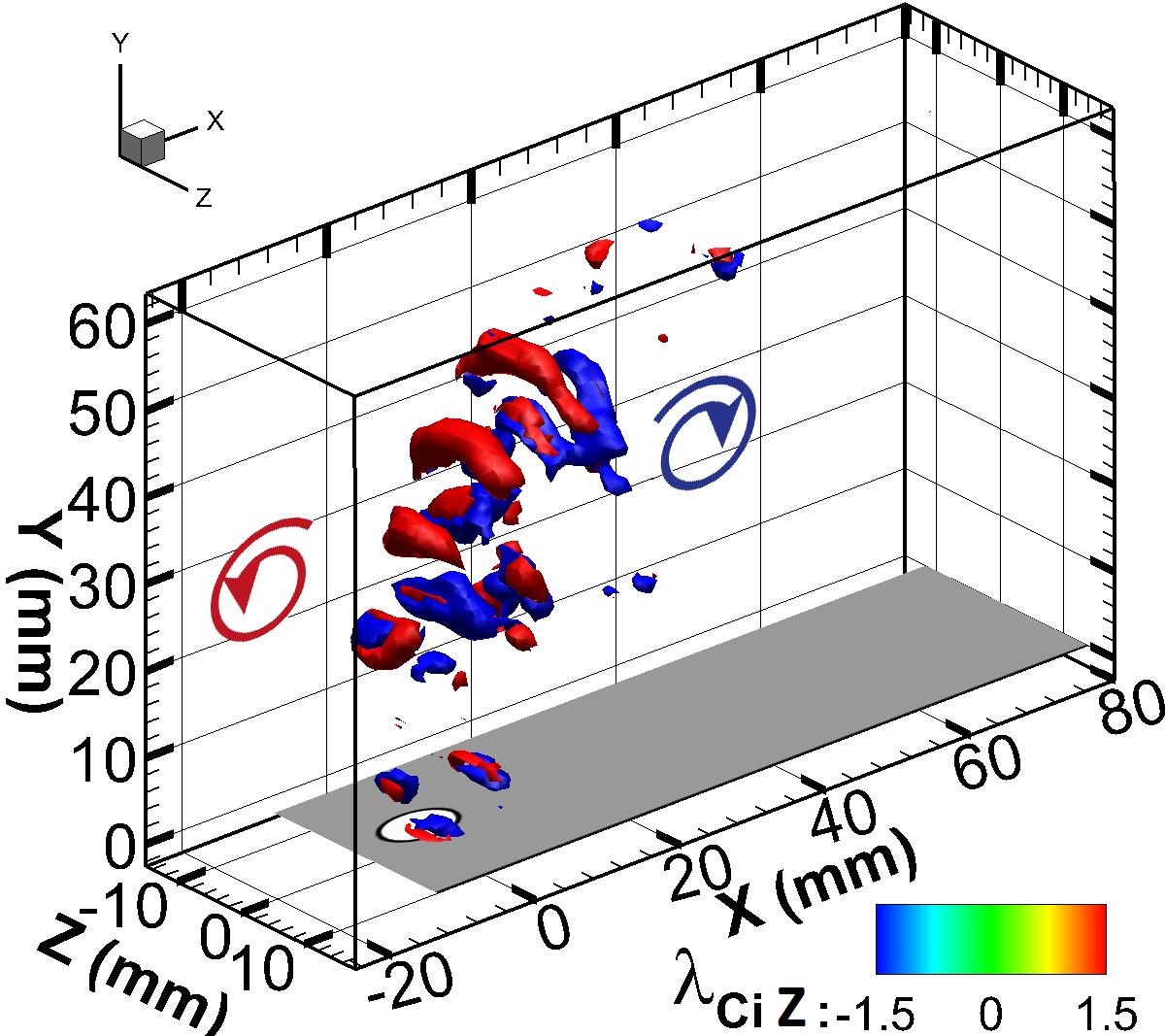} & 
	\includegraphics[width=0.4\textwidth]{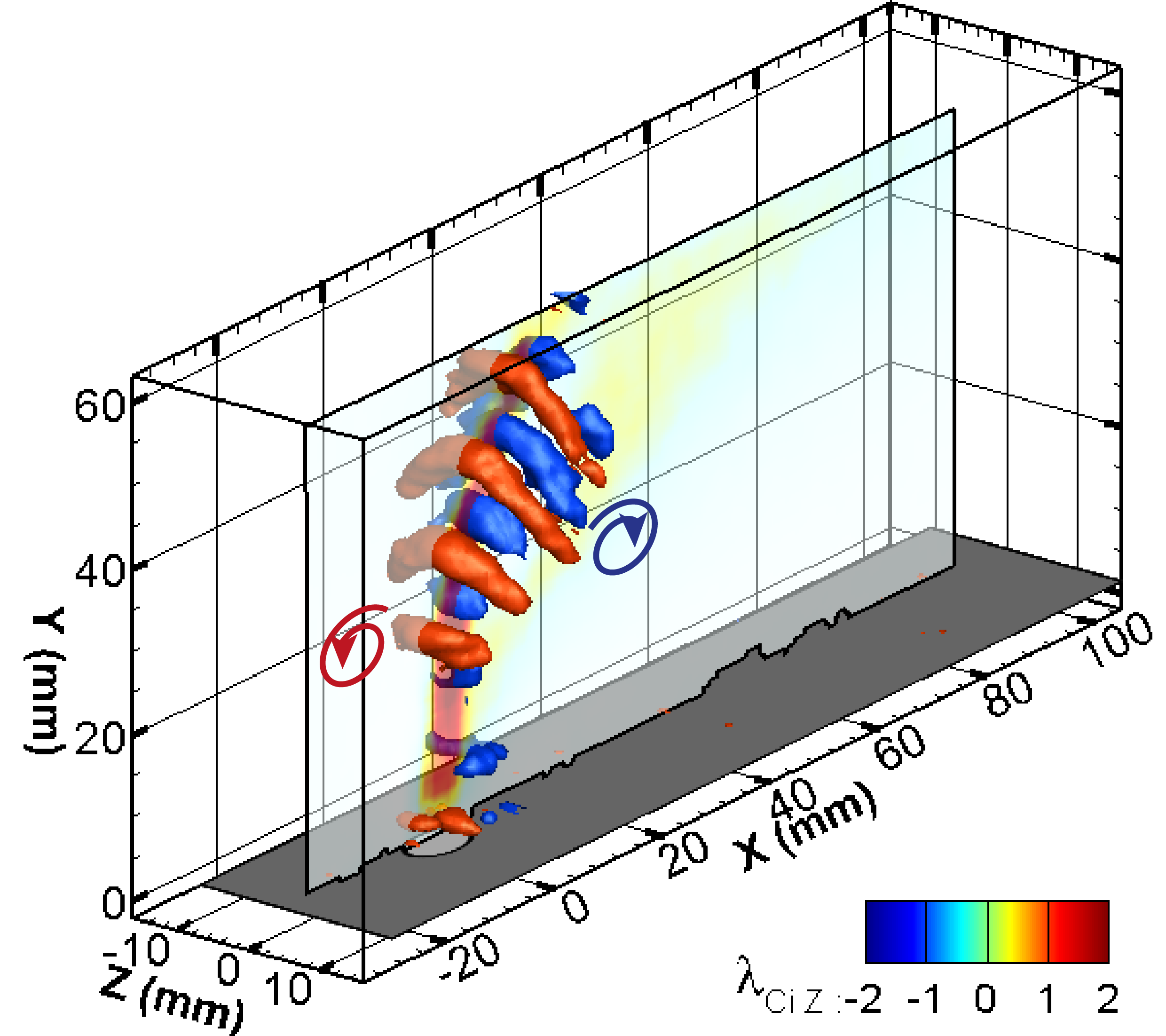}\\
a)&b)
\end{tabular}
\caption{a) Instantaneous velocity field. Isosurface of $\lambda_{ci\ Z}=\pm1.5\ s^{-1}$, b) Corresponding global conditional average. Isosurface of $\lambda_{ci\ Z}=\pm0.7\ s^{-1}$.}
\label{fig:MCGResults}
\end{center}
\end{figure}
If enough instantaneous fields are associated with a phase, not only we can have a statistically converged GCA, but it also becomes possible to improve the resolution of the interpolation grid. For example for the GCA in Fig. \ref{fig:MCGResults}b, the voxel interpolation size is 1 vector/mm. It is a slightly better resolution than the instantaneous field (Fig. \ref{fig:MCGResults}a) : 1 vector/1.2mm. Following \citet{CaAi14} work, this instantaneous resolution has been optimized to be the best possible, i.e the smallest resolution with at least 99\% of the voxels in the measurement volume successfully interpolated. In practice, there are to be at least 7 raw velocity vectors by voxel for the interpolation to be reliable\citep{Pereira2002}.  Therefore,  the greater the density of raw velocity vectors of the concatenated field is, the more resolved the interpolation can be.  Nevertheless, a compromise has to be found between the convergence of the GCA and the final resolution of the interpolated velocity field since smaller voxels contains less raw velocity vectors. For instance, 39 instantaneous fields have been used for the GCA in Fig. \ref{fig:MCGResults}b. A 17\% diminution of the voxel size corresponds to a 42\% diminution of the voxel volume : $1^3/1.2^3=0.58$. This ratio can also be seen as a diminution ratio of the concentration of raw velocity vectors per voxel. Statistically, the equivalent concentration corresponds to an average realized with $39\cdot0.58\approx23$ instantaneous fields. Obviously, like every time-average method, the resolution gain is directly correlated with the number of instantaneous fields associated with each phase. We get here a 17\% gain on the final resolution of the interpolated velocity field. However, with enough instantaneous fields, it is possible to obtain a final resolution smaller than the camera pixel size, only limited by the sub-pixelic accuracy of the particle detection algorithm\citep{Kahler2012}. Of course, 23 velocity fields are not enough to get a statistically converged time-average. Nevertheless, it is enough to have better visualizations and to perform a proof of principle. 
\section{Local Conditional Average (LCA)}
\subsection{Construction}
\begin{figure}[htbp!]
\begin{center}
\begin{tabular}{cc}
	\includegraphics[width=0.33\textwidth]{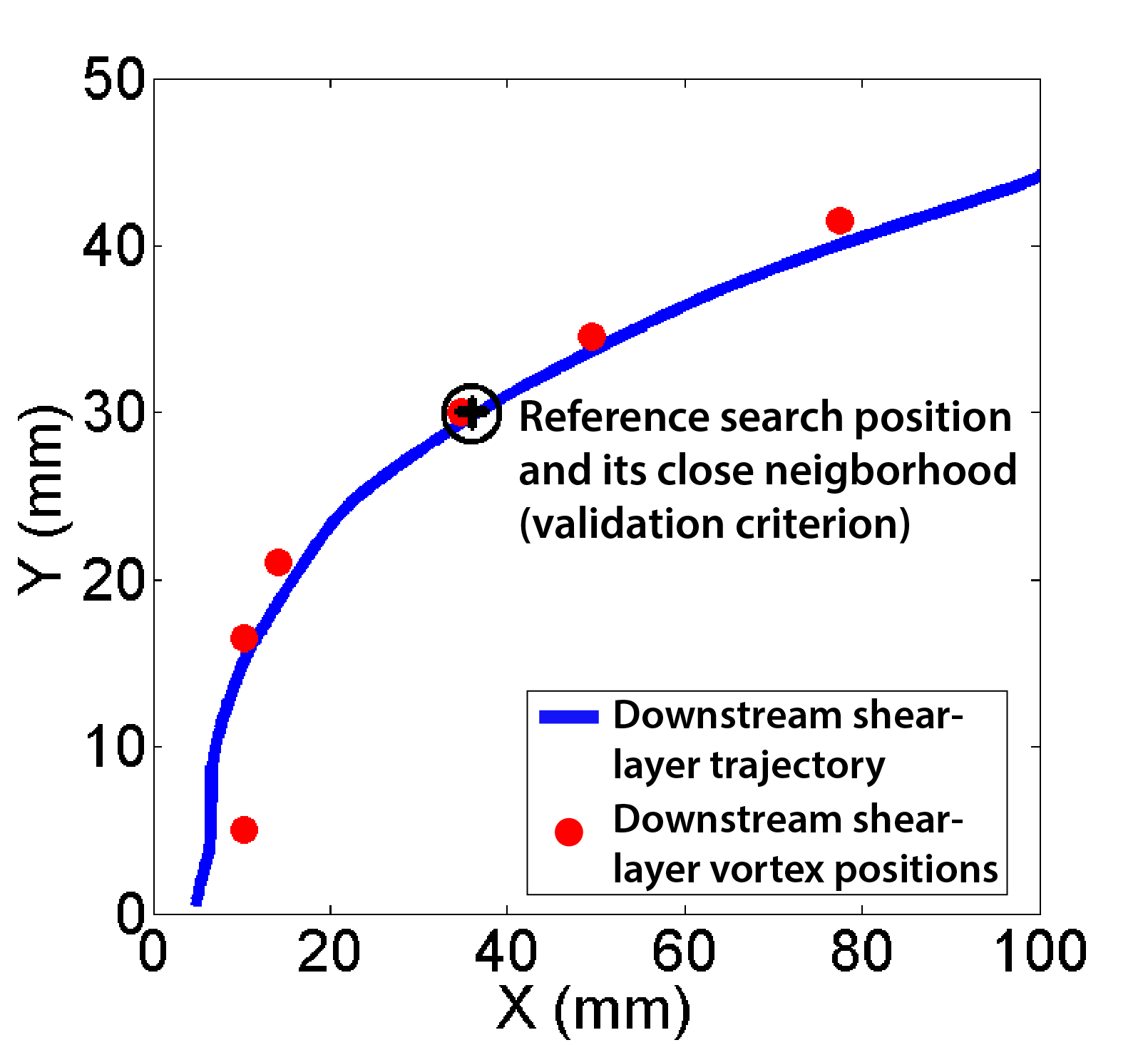} & 
	\includegraphics[width=0.5\textwidth]{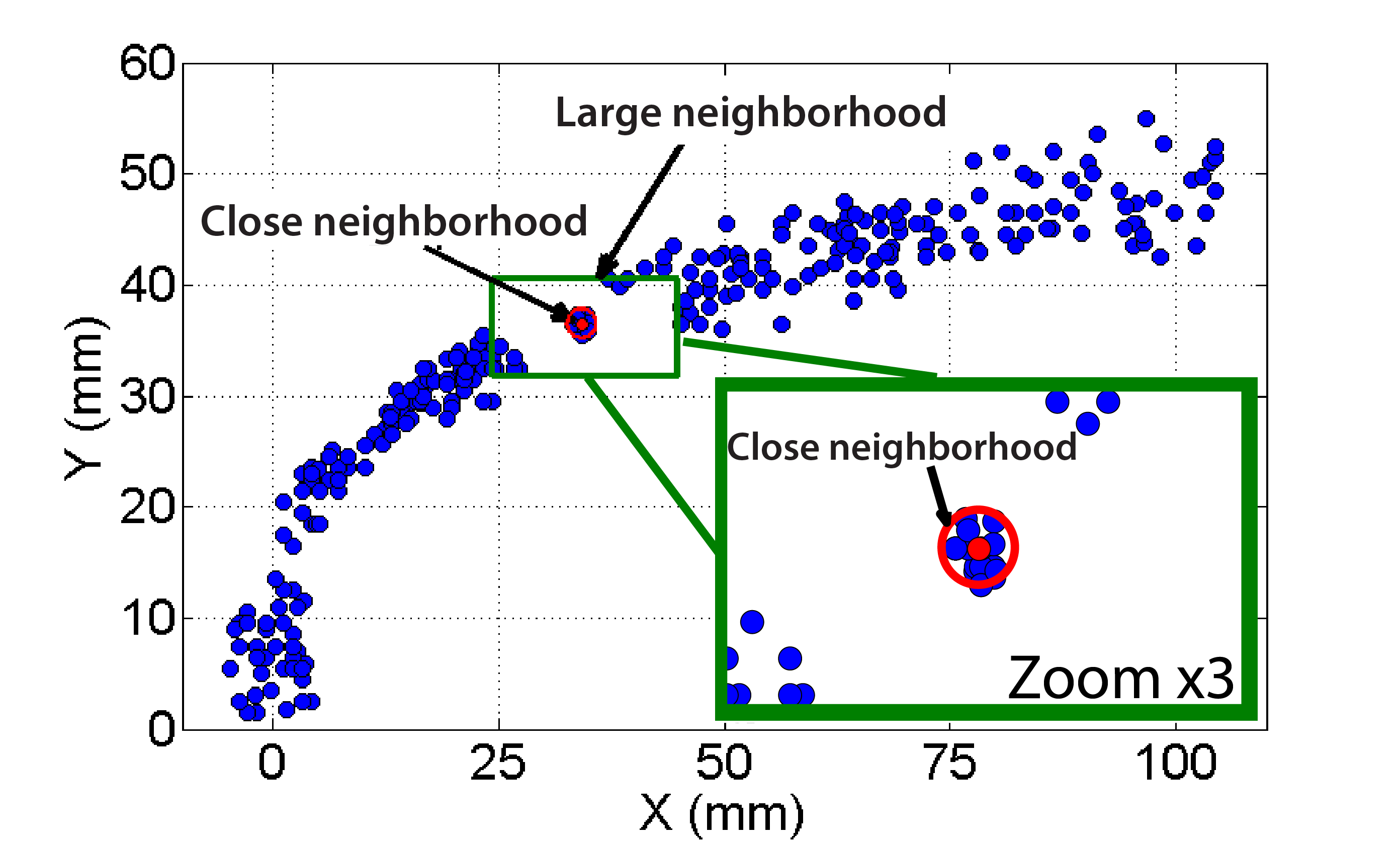}\\
a)&b)
\end{tabular}
\caption{a) Positions of the downstream shear-layer vortices detected in the symmetry plane of an instantaneous field which satisfy the validation criterion of the LCA : one of the vortices is inside the neighborhood of the reference search position. b) Cumulative view of all the positions of the downstream shear-layer vortices of the fields satisfying the validation criterion.}
\label{fig:MCL-construction}
\end{center}
\end{figure}
For the local conditional average, only one reference position is considered, therefore the LCA can be used everywhere in the measurement volume. 
In this article, the reference position is taken along the shear-layer trajectory for the sake of comparison between GCA and LCA. 
A close neighborhood around this position is used to sort the instantaneous velocity fields. 
For each instantaneous field, if one of the shear-layer vortices is included in the close neighborhood, the instantaneous field is kept for the LCA (Fig. \ref{fig:MCL-construction}a). Fig. \ref{fig:MCL-construction}b shows a cumulative view of the vortices positions of the kept instantaneous fields. In the same way as the GCA, all the valid raw velocity fields are then concatenated into a single dense raw velocity field.  Finally, an interpolation of this dense raw velocity field on a 3-dimensional grid complete the LCA.
Figure \ref{fig:MCL-Results}c shows the resulting LCA processed from 83 instantaneous velocity fields one of which is shown in Fig. \ref{fig:MCL-Results}a. The position of the LCA volume is also shown in Fig. \ref{fig:MCL-Results}a and a zoomed view on this volume is given in Fig. \ref{fig:MCL-Results}b to make the comparison easier. While the hairpin vortex is noisy and difficult to discern in the instantaneous field (Fig. \ref{fig:MCL-Results}b)%
, it is efficiently recovered with the LCA (Fig. \ref{fig:MCL-Results}c).
\subsection{Results and discussion on LCA}
\begin{figure}[htbp!]
\begin{center}
\begin{tabular}{ccc}
	\includegraphics[width=0.25\textwidth]{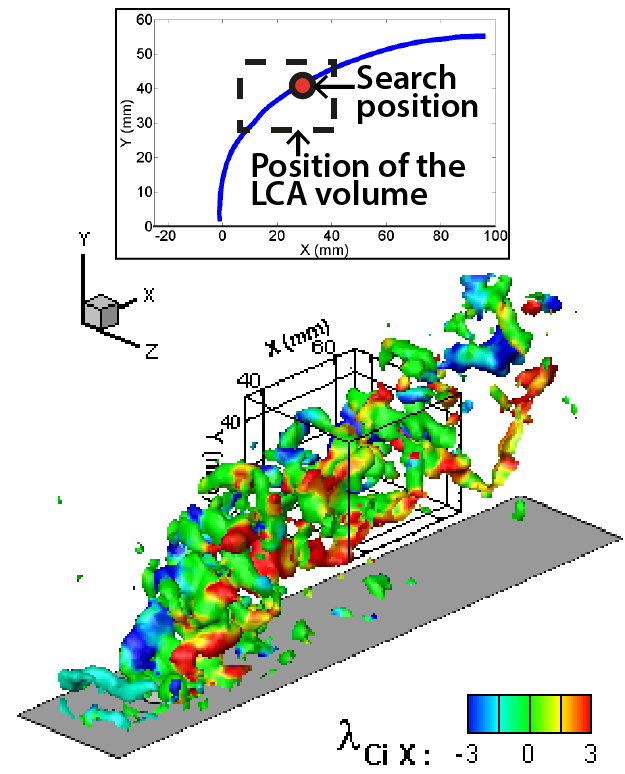} & 
\includegraphics[width=0.33\textwidth]{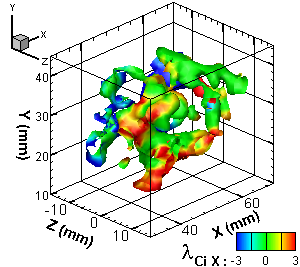} & 
	\includegraphics[width=0.35\textwidth]{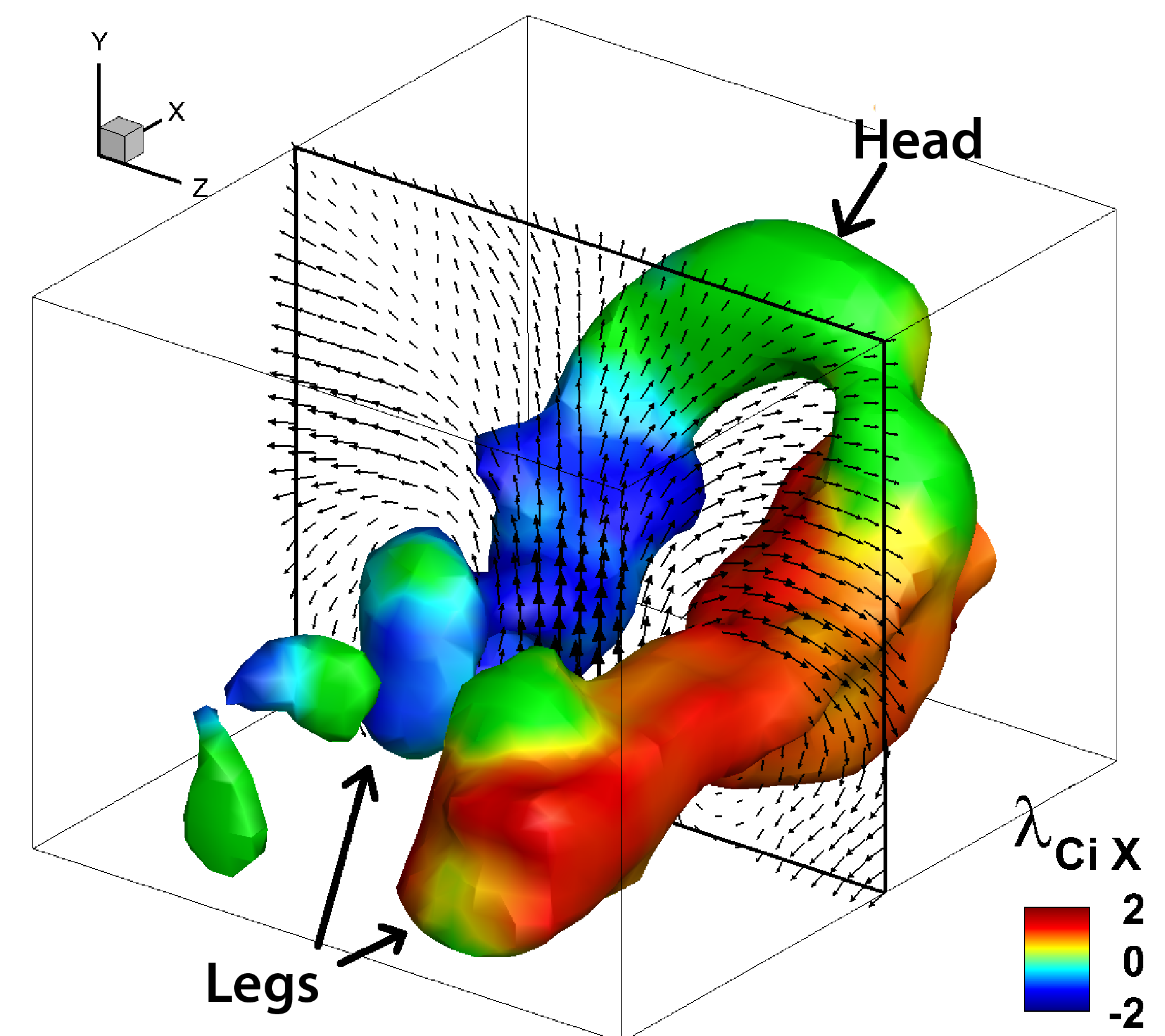}\\
a)&b)&c)
\end{tabular}
\caption{Isosurface of $|\lambda_{ci}|=2.7\ s^{-1}$ colored in $\lambda_{ci\ X}$ : a) Complete view of one of the instantaneous fields used in the LCA of figure c. Locations of the LCA volume and of the search position. b) Zoomed view on the LCA volume (1.2 vec/mm). c) Corresponding LCA (0.75 vec/mm). Isosurface of $|\lambda_{ci}|=2\ s^{-1}$ colored in $\lambda_{ci\ X}$.}
\label{fig:MCL-Results}
\end{center}
\end{figure}
Since the condition of the local conditional average is less restrictive (only one reference position), a much smaller neighborhood has been taken. 
The neighborhood radius for the local method is 0.5-mm long (Fig. \ref{fig:MCL-construction}) against 2 mm for the global method (Fig. \ref{fig:MCG-construction}). 
As a result, near the reference position the LCA, the reconstructed values of the different fields (velocity, vorticity, swirling fields) are closer from the instantaneous measured values than the reconstructed values for GCA. 
For instance, using the LCA, the maximal swirling intensities of both the head ($\lambda_{ci\ Z}\approx1.3\ s^{-1}$) and the legs ($|\lambda_{ci\ X}|\approx3.18\ s^{-1}$) of the hairpin vortices are found to be quantitatively much closer to the ones observed in the instantaneous fields (resp. $\lambda_{ci\ Z}=1.65\ s^{-1}$, $|\lambda_{ci\ X}|=2.3\ s^{-1}$) than the swirling intensities of the hairpin vortex obtained by GCA at the same position (resp. $\lambda_{ci\ Z}\approx0.9\ s^{-1}$, $|\lambda_{ci\ X}|\approx1.3\ s^{-1}$). 
The experimental values have been determined as the mean values of the distributions  of  $\lambda_{ci\ Z}$ and $|\lambda_{ci\ X}|$ instantaneous swirling intensities (generated from the 1000 instantaneous fields) at the same position.\\
Moreover, because the research neighborhood radius of the LCA is a much less restrictive condition than the one used for GCA, a bigger number of instantaneous fields satisfy the condition and are kept : 83 instantaneous fields for the LCA in Fig. \ref{fig:MCL-construction}b and \ref{fig:MCL-Results}c versus 39 for the GCA. 
It makes possible to increase the LCA vector field resolution while controlling the effective number of instantaneous field used for the local averages inside each voxels.
For example, in our experiments the instantaneous velocity fields (Fig. \ref{fig:MCL-Results}a,b for instance) have a resolution of 1 velocity vector/1.2 mm. This resolution of interpolation is the best one achievable on this setup taking into account the local particle concentration in the water, the screening effect of particles between each other, and the conditions to have adequately and quasi-entirely interpolated ($>$99\%) velocity fields \citep{CaAi14}. By keeping the same resolution (1.2 vec/mm), the voxels keep the same volume than the volume of the instantaneous voxels. With 83 instantaneous fields, the concentration of raw vectors by voxel is then multiplied by 83. By increasing the resolution (0.75 vec/mm in Fig. \ref{fig:MCL-Results}c), the voxel volume is decreased. As a result the concentration of raw vectors by voxel gets smaller. In the  Fig. \ref{fig:MCL-Results}c case, the volume reduction is $0.75^3/1.2^3=0.24$. One can then consider that the equivalent local average in each voxel is performed on approximately $0.24\times0.83=20$ fields. Once again, 20 velocity fields are not enough to get a statistically converged time-average. Nevertheless, it is enough to have better visualizations and to perform a proof of principle. From 1.2 to 0.75 vec/mm, a 38\% gain on the final resolution of the interpolated velocity field is obtained. 
Of course, the LCA is only relevant in a limited volume around the reference position (large neighborhood in Fig. \ref{fig:MCL-construction}b : 3$\times$3$\times$3 $cm^3$). Outside the close neighborhood, a degradation is therefore expected and the results have to be taken more qualitatively.  The loss of spatial correlation with the distance to the reference position has not been extensively quantified. In fact, it would be meaningless. Indeed, rather than determine the limitations of the local conditional averaging process, it would quantify the dispersion of the topological forms taken by the vortices which are issued from the close neighborhood. Nevertheless, the relative errors on the swirling intensities of the head and legs of the hairpin vortices make possible to estimate this degradation.  The relative error on the swirling intensities of the head is $100\times(\lambda_{ci\ Z\ LCA}-\lambda_{ci\ Z\ measured})/\lambda_{ci\ Z\ measured}=21.2\%$, for the legs the relative error is $100\times(\lambda_{ci\ X\ LCA}-\lambda_{ci\ X\ measured})/\lambda_{ci\ X\ measured}=27.7\%$ . The relative closeness of these values shows that the 83 vortices in the close neighborhood are all hairpin vortices with a similar shape. The closeness of the dispersion of the hairpin\rq{}s legs and heads is a good indicator to validate this LCA method. In the case of the quasi-periodic vortices induced by a jet in crossflow, the LCA is therefore a powerful tool to identify and visualize the instantaneous swirling structures.
\section{Conclusion}
Because most optical volumetric velocimetry techniques use tracers to measure the flow velocity field, the achievable tracer concentration is inherently limited by the optical screening of the tracers. This phenomenon induces an optimal resolution of the final velocity fields\citep{CaAi14}. As a result, while it is theoretically always possible to increase the resolution of numerical velocity fields to observe smaller flow structures (for a computational cost), the resolution of instantaneous 3D velocity fields is intrinsically limited. 
To overcome this limitation, 2 different conditional averaging processes have been developed and are described in this study. They have been developed from experimental 3D velocity field of a jet in crossflow, a complex flow configuration where multiples vortical structures interact, and whose pseudo-periodicity makes it indispensable to use these conditional averages rather than a classical phase average. The global conditional average (GCA) uses the spatial distribution of the vortices in the symmetry plane to extract statistical properties of the vortices, to quantify the jet pseudo-periodicity, and finally reconstruct pseudo-periodically corrected phases. On the opposite, the local conditional average (LCA) focuses on one specific reference position in the flow field. More than a corrected phase average, the LCA is rather a state average, combining the different 3-dimensional forms of the vortices passing by the neighborhood of the reference position.
Both conditional average makes possible to retrieve the main features of the experimental instantaneous fields, while significantly reducing the experimental noise and increasing the vector field resolution. For the specific example shown in this article we obtain a 17 \% gain for the GCA and 38\% gain for the LCA. With enough instantaneous fields, it becomes possible to obtain statistically converged conditional averages and a sub-pixelic resolution of the final vector field.  
Both methods have been applied to a jet in crossflow configuration but could be easily implemented and used on other flows, as long as a statistical analysis of the flow periodicity or pseudo-periodicity can be achieved.

\bibliographystyle{IEEEtranSN}
\bibliography{LowVelocityRatioTopology2}
\end{document}